\documentclass[twocolumn]{aastex62}
\usepackage{amsmath}
\usepackage[normalem]{ulem}
\usepackage{bm}
\usepackage{comment}
\newcommand{\msun}{M$_\odot$}
\newcommand{\mdot}{\dot{\mathcal{M}}}

\newcommand{\qam}[1]{{\color{blue} #1}}

\newcommand{\reduction}{\sim30}

\shorttitle{}
\shortauthors{Mabanta et al.}

\begin{document}

\title{Convection-Aided Explosions in One-Dimensional Core-Collapse
  Supernova Simulations I: Technique and Validation}

\correspondingauthor{Quintin A.~Mabanta}
\email{qam13b@my.fsu.edu}

\author{Quintin A.~Mabanta}
\affiliation{Physics, Florida State University, Tallahassee, FL, USA}
\affiliation{Los Alamos National Laboratory}

\author{Jeremiah W.~Murphy}
\affiliation{Physics, Florida State University, Tallahassee, FL, USA}

\author{Joshua C.~Dolence}
\affiliation{Los Alamos National Laboratory}

\begin{abstract}
Most one-dimensional core-collapse simulations fail to explode,
  yet multi-dimensional simulations often explode.  A dominant
  multi-dimensional effect aiding explosion is neutrino-driven convection.
We incorporate a convection model in approximate one-dimensional core-collapse
  supernova (CCSN) simulations. This is the 1D+ method.  This convection model lowers the
  neutrino luminosity required for explosion by $\reduction$\%, similar to
  the reduction observed in multi-dimensional simulations. 
The model is based upon the global
  turbulence model of \citet{mabanta18} and models the
  mean-field turbulent flow of neutrino-driven convection.  
  In this preliminary investigation, we use simple neutrino heating
  and cooling algorithms to compare the critical condition in the 1D+
  simulations with the critical condition observed in two-dimensional
  simulations.  Qualitatively, the critical conditions in the 1D+
  and the two-dimensional simulations are similar.  The assumptions
 in the convection model affect the radial profiles of
density, entropy, and temperature, and
comparisons with the profiles of
three dimensional simulations will help to calibrate these
  assumptions. These 1D+
simulations are consistent with the profiles and explosion conditions of
equivalent two-dimensional CCSN
simulations but are $\sim$10$^2$ times faster, and the 1D+ prescription has the potential to be
  $\sim$10$^5$ faster than three-dimensional CCSN
simulations.  The 1D+ technique
will be ideally suited to test the explodability of thousands of
progenitor models.
\end{abstract}

\keywords{supernovae: general --- hydrodynamics --- methods: analytical
  --- methods: simulation --- shock waves --- turbulence}

\section{Introduction}
\label{sec:introduction}

Progress in understanding the core-collapse problem has
  required three general approaches.  One approach is to simulate
  core-collapse supernovae using multi-dimensional, radiation
  hydrodynamic simulations.  On	the one hand, numerical simulations will provide quantitative
	predictions for CCSN explosions, on the other hand these require
	considerable computational effort.  A recent simulation of a
	successful explosion required $\sim$18 million CPU-hours
	\citep{vart18}.  On
	16,000 cores this would take roughly 1.5 months of nonstop
	computing. Another approach is to investigate analytics.
  Analytic investigations such as  \citet{bethe90,burrows93,thompson00,janka01,pejcha12,murphy17} provide a
	deeper understanding of the explosion conditions, but  at the cost of quantitative
  accuracy.
Another approach is to incorporate multi-dimensional
  effects in one-dimensional simulations.  In
this manuscript, we adopt this latter approach, implementing a
turbulence model in the one-dimensional rendition of FORNAX for a 13 $M_\odot$ progenitor from \citet{wh07}. Additionally, we develop the
technique (1D+) and show that it reproduces the reduction in the critical
curve observed in multi-dimensional simulations.

Of the many attempts to quantify a critical condition for
explosion \citep{burrows93,thompson00,janka01,pejcha15,murphy17}, the
one proposed by \citet{burrows93} has proven quite useful in
  comparing the explosion conditions of simple simulations.  
	   In solving for stalled
	  accretion shock solutions, they found no stalled solutions above
	  a critical curve in neutrino luminosity and mass accretion
	  rate.  They suggested but did not prove that the solutions above
	  this curve are explosive.  \citet{murphy17} analyzed the
	  solutions above this curve and found them to have positive shock
	  velocity, once again suggesting explosive solutions.
	  Furthermore, \citet{murphy17} expanded the critical condition
	  to a critical hypersurface among five parameters:  mass accretion rate ($\mdot$), neutron star mass ($M_{NS}$), neutron star radius ($R_{NS}$), and neutrino temperature ($T_\nu$).

This critical condition has been useful in quantifying and
  explaining the impact
of turbulence on explosion outcomes.  For example, \citet{murphy08b}
found that the critical neutrino luminosity is a viable explosion
condition for 1D and 2D simulations, and they found that the critical
condition is 30\% lower for 2D simulations.  Many have since confirmed
these results in other multi-dimensional simulations  \citep{hanke12,dolence13,couch13}.  Some simulations suggest a slight difference between 2D and 3D, but detailed analysis of the critical condition shows that the difference between 2D and 3D is modest ($\lesssim$ 5\%) compared to the significant drop
(30\%) in going from 1D to multi-D \citep{hanke12,dolence13,handy14,fernandez15}.  Many numerical investigations \citep{herant94,janka95,burrows95,janka96,burrows07b,melson15,dolence15,muller16,roberts16,bruenn16} 
strongly suggested but did not prove that turbulence is responsible
for this reduction.

\citet{mabanta18} used the critical conditions to
  investigate if turbulence could reduce the explosion condition and
  how.  They incorporated a neutrino-driven
convection model \citep{murphy11,murphy13} into the critical condition analyses of
\citet{burrows93} and \citet{murphy17}.  They found that this
neutrino-driven convection model indeed reduces the critical condition by about 30\%.  Furthermore, they isolated the dominant turbulent
terms and quantified how each of these terms reduces the condition.
They found that turbulent ram pressure has some effect, but the
turbulent dissipation accounts for more than half of the reductions in
the critical condition.

Another approach to exploring which stars cross this critical condition is to force
one-dimensional simulations to explode with similar outcomes as
either nature or simulations. The first to pursue this was
\citet{ugliano12}; they removed the proto-neutron star from their 1D
simulations and replaced it with an inner boundary that contracted
with time, emulating the contraction due to neutrino
cooling. They calibrated this technique using observational
  constraints from SN~1987A.  Then, they examined a large suite of progenitor models and artificially triggered supernovae. Furthermore, \citet{perego15} performed a similar study which developed a generalized method to produce nucleosynthesis yields, neutron-star remnant masses, and explosion energies for several progenitors. This technique, coined PUSH, has the benefit in that 1D simulations are orders of magnitude less computer intensive  than their multi-dimensional counter-parts. Once again, their results depend upon the calibration of their parameters.  Moreover, forced explosions generally do not capture the
  multi-dimensional turbulence that aides explosion.  Therefore, it is
  not clear if they actually mimic the explosion conditions in
  simulations or nature.  None-the-less, such explorations have
  already shed light on new potential explosion outcomes.  For
  example, many authors have further noted that explosion outcome may not be
  monotonic with progenitor mass.  In particular, \citet{sukhbold16} note
  that stars below about 15 $M_\odot$ generically explode by these 1D studies,
  but between 21 $M_\odot$ and 25 $M_\odot$ they rarely explode, and above 27 $M_\odot$ there are
  islands of explodability.  However, the qualitative outcome of these
  1D studies may depend upon the nature of forced-explosion algorithms.

Multi-dimensional simulations self-consistently include all
  effects (neutrinos and multi-dimensional instabilities) that aid
  explosion.  However, these simulations are expensive and it may not
  be feasible to properly explore the statistics of explodability.  For example,
  \citet{sukhbold18} recently noted that the Fe-core mass is not monotonic
  with progenitor mass.  E.g., the Fe-core mass for a 15 M$_{\odot}$
  progenitor is 1.580 M$_\odot$, while the Fe-core mass for a 15.01 M$_{\odot}$ progenitor is 1.513 M$_\odot$. This may imply that  the ease of explodability is not monotonic either.  Therefore, to
  predict which stars explode may require hundreds, if not thousands, of
  simulations.  At the current rate of simulating one multi-dimensional
  model every 1.5 months, a full systematic exploration of
  explodability using 3D simulations would take nearly a millennium.  On the
  other hand, if one could mimic the explosion conditions of multi-dimensional
  simulations in one-dimensional simulations, then this systematic study would
  take only a few weeks.


Here, we include a neutrino-driven convection model into
  one-dimensional simulations.  These 1D+ simulations promise to
  capture the explosion conditions of multi-dimensional simulations,
  but remain orders of magnitude faster.  The algorithms in this
  manuscript are an extension of the techniques employed in \citet{mabanta18}. To test the validity of
the 1D+ algorithm, we compare the critical condition in 1D, 1D+,
  and 2D simulations.  Hence, we incorporate a turbulence model in
  one-dimensional simulations to explore explodability.


In section \ref{methods}, we describe the technique for incorporating the
  turbulence model in a one-dimensional radiation hydrodynamics code.
In section \ref{results}, we test the technique using a simple light bulb
model, explore how the model affects the profiles, and thus how the critical condition for
explosion is modified. We also discuss how comparing these one-dimensional
profiles with multi-dimensional profiles will constrain the effects and form of
the turbulence model.  Finally, we conclude in section \ref{conclusion} that our results are a valid approximate substitute for a multi-dimensional analysis in the context of the critical curve.

\section{Methods}
\label{methods}

Fundamentally, the 1D+ method is a one-dimensional hydrodynamics
  method with a mean-field turbulence model.  The turbulence model
  captures the dominant mean-field characteristics of neutrino-driven
  convection without the need of simulating a full three-dimensional
  simulation.  The first step in modifying a one-dimensional code is
  identifying the new turbulent terms.  In the following sections, we
  use Reynolds decomposition to derive the evolution equations for the background flow including
  turbulent terms.  Then we present two turbulence models.  One is for
  the gain region, and the other is for protoneutron star (PNS)
  convection.  Finally, we present the numerical techniques for
  implementing these turbulence models into one-dimensional simulations
of FORNAX.

\subsection{Reynolds Decomposed Equations for Spherical Symmetry}
The governing conservation equations are:
\begin{equation}
\rho_{,t} + (\rho u^i)_{;i} = 0 \, ,
\label{mass}
\end{equation}
\begin{equation}
(\rho u_i)_{,t} + (\rho u_i u^j + \delta^j_i P)_{;j} =  - \rho \Phi_{,i}  \, ,
\end{equation}
and
\begin{equation}
(\rho E)_{,t} + \left [\rho u^j \left (h + \frac{u^2}{2} \right ) \right ]_{;j} = - \rho u^j \Phi_{,j} + \rho q \, .
\label{energy}
\end{equation}
Where $\rho$ is mass density, $u$ is velocity, $P$ is
  pressure, $\Phi$ is gravitational potential, $h$ is enthalpy, and $q$ is
  the total heating. In general, heating and cooling by
neutrinos is best described by neutrino transport  \citep{janka07,janka17a,tamborra17};
in this first test of the convection model, we invoke a
simple light-bulb prescription for neutrino heating and cooling \citep{janka01}
\begin{equation}  
\begin{aligned}    
q = H_0 \left ( \frac{\textrm{$10^7$ cm}}{r}\right )^2 \left ( \frac{L_\nu}{\textrm{$10^{52}$ ergs}}\right ) \left ( \frac{T_\nu}{\textrm{4 MeV}}\right )^2 \\ - C_0 \left (\frac{T}{\textrm{2 MeV}}\right )^6 \, .
\end{aligned}
\end{equation}
$L_\nu$ is the neutrino luminosity emitted from the core of the star, $T$ is the matter temperature, $T_\nu$ is the neutrino temperature, $H_0$ is the heating factor ($1.544 \times 10^{20}$ ergs/g/s) and $C_0$ is the cooling factor ($1.399 \times
  10^{20}$ ergs/g/s).  See \citet{janka01} for details.

Next, we Reynolds decompose the flow variables into background variables (0
subscript) and turbulent variables ($\prime$ superscript).  For
example, the Reynolds decomposed density is $\rho = \rho_0 +
\rho^{\prime}$.  We then
insert these decomposed variables into the hydrodynamics equations
and average over solid angle and a small window in time.  In the
  following discussion, this average is represented by $\langle \cdot
  \rangle$.  The
resulting equations describe the evolution of the spherically
symmetric background flow and
self-consistently include turbulent correlations.  The full decomposed equations are 
\begin{equation}
\rho_{0,t} + (\rho_0 u^i_0)_{;i} + \langle \rho' u^{i\prime} \rangle_{;i} = 0
\label{consmasss}
\end{equation}
\begin{equation}
\begin{aligned}
( \rho_0 u_{0i} + \langle \rho' u^{\prime}_i \rangle )_{,t} +  \left [ P_0\delta^j_{i} + \rho_0u_{0i}u^j_0 \right ]_{;j} + \rho_0 \Phi_{,i} \\ = - \left [ \langle \rho R_i^j \rangle + u_{0i} \langle \rho'u^{j\prime} \rangle + u_{0j} \langle \rho'u^{i\prime} \rangle \right ]_{;j} 
\label{consmom}
\end{aligned}
\end{equation}
\begin{equation}
\begin{aligned}
\langle \rho E \rangle_{,t} + \langle \rho E u^i_0 + u^i_0P_0 \rangle_{;i} + \rho_0u^i_0\Phi_{,i} - \rho_0 q \\ = - \langle F^i_P + F^i_I + F^i_K - u^{j \prime} \sigma^{i \prime}_{j}   \rangle_{;i} + W_b \, ,
\label{consenergyy}
\end{aligned}
\end{equation}
where $F^i_P = \rho P u^{i \prime}$ is the perturbed pressure flux, $F^i_I = \rho e u^{i \prime}$ is the perturbed internal energy flux, $F^i_K = \rho u^{\prime 2}u^{i \prime }$ is the kinetic energy flux, and $W_b = \rho'u^{i\prime}g_i$ is the work done by buoyant driving. 
Many of these turbulent correlations are negligible for the context of
neutrino-driven convection.  If one assumes steady state for the
turbulent correlations, then
\begin{equation}
\langle \rho E \rangle_{,t} = (\rho_0e_0 + \frac{1}{2}\rho_0u_0^2)_{,t} \, .
\end{equation}
Additionally, if we define the ram pressure as $\rho R^j_i = \rho u_iu^{j\prime}$, Equation~\ref{consmom} becomes:
\begin{equation}
\begin{aligned}
( \rho_0 u_{0i} )_{,t} +  \left [ P_0\delta^j_{i} + \rho_0u_{0i}u^j_0 \right ]_{;j} + \rho_0 \Phi_{,i} = - \langle \rho R_i^j \rangle_{;j} \, . 
\end{aligned}
\end{equation}
Under the Boussinesq approximation, the buoyant correlation is larger
than correlations involving pressure perturbations.  Therefore, we
ignore $F^i_P$.  Since $P \sim \rho e$, this approximation also means
that correlations just involving $\langle \rho^\prime e^\prime \rangle$ are also
negligible.  This reduces the total energy flux term to
\begin{equation}
\langle \rho E u^i_0 \rangle_{;i} = (\rho_0e_0u^i_0)_{;i} + \frac{1}{2}( \rho_0u_0^2u^i_0 + \rho_0u^i_0\langle u^{\prime 2} \rangle)_{;i} \, .
\end{equation}
$F^i_K$ is a third order term for velocity that is generally
  found to be small (see \citet{murphy11}).
  For a more thorough discussion of which terms may
be combined or ignored see \citet{canuto93} and \citet{murphy13}.  After employing the
assumptions above, the Reynolds decomposed
equations become
\begin{equation}
\rho_{0,t} + (\rho_0 u^i_0)_{;i} = 0  \, ,
\label{turbmass}
\end{equation}
\begin{equation}
( \rho_0 u_{0i})_{,t} +  \left [ P_0\delta^j_{i} + \rho_0u_{0i}u^j_0 \right ]_{;j} + \rho_0 \Phi_{,i} = - \langle \rho_0 R_i^j \rangle_{;j} 
\label{turbmom}
\end{equation}
and
\begin{equation}
\begin{aligned}
(\rho_0e_0 + \frac{1}{2}\rho_0u_0^2)_{,t} + (\rho_0e_0u^i_0)_{;i} + \frac{1}{2}( \rho_0u_0^2u^i_0)_{;i} \\ + \langle u^i_0P_0 \rangle_{;i} + \rho_0u^i_0\Phi_{,i} - \rho_0 q = - \langle F^i_I \rangle_{;i} + \langle W_b \rangle   \, .
\label{turbenergy}
\end{aligned}
\end{equation}
The terms on the left are those that are included in any standard
spherically symmetric hydrodynamics code.  The terms on the right are
new, and represent the effects of turbulence.

\subsection{Turbulence Model for the Gain Region}

Equations \ref{turbmass}-\ref{turbenergy} represent 3 evolution equations for 9 total unknowns. Choosing an equation of state (here, we use the SFHo \citep{hempel13}) reduces this number to 8.
  Since there are still more variables than equations, one must develop a
  turbulence model to close these equations.  \citet{murphy11} and \citet{murphy13}
  proposed a neutrino-driven turbulence model.  \citet{mabanta18} refined this
  model, included it in steady state equations and noted that it
  reduces the critical condition for explosion in accordance with
  multi-dimensional equations.  A reiteration of the turbulence model
  follows.

There are five turbulent variables ($\textbf{R},L_e,W_b$),
three of them are Reynolds stress 
  terms ($R_{rr}$, $R_{\phi \phi}$, and $R_{\theta \theta}$);
our five global constraints are as follows. First,
    we eliminate the tangential components of the Reynolds stress.  In neutrino-driven convection, there is a preferred direction (i.e. in the direction of gravity) and simulations show that there is an
equipartition between the radial direction and both of the tangential
directions \citep{murphy13}:
\begin{equation}
R_{rr} \sim R_{\phi \phi} + R_{\theta \theta} \, .
\label{firstr}
\end{equation}
Similar simulations showed that the transverse components are roughly the same scale:
\begin{equation}
  R_{\phi \phi} \sim R_{\theta \theta}  \, .
  \label{secondr}
\end{equation}
From \citet{murphy11}, we note that buoyant driving roughly
  balances turbulent dissipation:
\begin{equation}
W_b \approx E_k \, ,
\label{third}
\end{equation}
where the buoyant driving is the total work done by buoyant
  forces in the convective region,
\begin{equation}
W_b  = \int\limits_{r_g}^{r_s} \langle \rho' u'_i \rangle g^i dV \, ,
\label{wb}
\end{equation}
and the total power of dissipated turbulent energy is
\begin{equation}
\label{eq:Ekdefine}
E_k = \int\limits_{r_g}^{r_s} \rho \epsilon_k dV \, .
\end{equation}
Lastly, both two- and three-dimensional simulations from
\citet{murphy13} show that the 
the neutrino power absorbed in the gain region is related to the turbulent luminosity and the turbulent dissipation by
\begin{equation}
L_e^{max} = \alpha L_\nu \tau
\label{tba}
\end{equation}
and
\begin{equation}
E_k = \beta L_\nu \tau   \, ,
\label{first}
\end{equation}
where simulations found these values to be $\alpha \approx$ .55 and
$\beta \approx$ .3 for two-dimensional simulations, and $\alpha
  \approx .7$ and $\beta \approx .3$ for three-dimensional
  simulations.  Together, equations~(\ref{firstr}-\ref{first}) represent our turbulence closure model.

Equations~(\ref{third}-\ref{first}) represent a global turbulence model, but the
  evolution equations require local turbulent terms.
To
  translate the global model into a local model, we make assumptions about the radial
profile for each term.  To ensure that the turbulent profiles
  satisfy the global conditions, each turbulent profile is scaled by a
scale factor.   
We introduce three scale factors for the turbulent region: a constant Reynolds stress
($\textbf{R}$), a constant dissipation rate ($\epsilon_k$), and a
maximum for the turbulent luminosity ($L_e^{max}$); the corresponding
local terms are $\nabla \cdot \langle \rho \textbf{R} \rangle$,
$\langle W_b \rangle$, and $\nabla \cdot \langle \vec{F}_e \rangle$ respectively (see equations (\ref{turbmom}-\ref{turbenergy})). Thus, the final solution for turbulence boils down to finding these three parameters.

Kolmogorov's theory of turbulence predicts the turbulent dissipation rate scales as the
perturbed velocity cubed over the characteristic length of the
instabilities \citep{kolm}. Numerical simulations suggest
  that the scale of this length in convection is roughly the size of
  the convective zone  \citep{murphy11,murphy13,couch14,fogl15,fern15}, or the gain region in the core-collapse case. Hence, we relate the Reynolds stress
to the turbulent dissipation by: 
\begin{equation}
\epsilon_k \approx \frac{u'^3}{\mathcal{L}} \ =
\frac{R_{rr}^{3/2}}{\mathcal{L}} \, ,
\label{second}
\end{equation}
where $\mathcal{L}$ is the largest turbulent eddy size. 
  We assume that
$\epsilon_k$ is constant over the gain region. Therefore, from equation~(\ref{eq:Ekdefine}),
\begin{equation}
\label{Ek}
  E_k \approx \epsilon_k\int_{r_g}^{r_s} \rho  dV \, ,
\end{equation}
we have
\begin{equation}
  \epsilon_k \approx \frac{E_k}{M_{gain}} = \frac{W_b}{M_{gain}} \, .
  \label{epk}
\end{equation}
Finally, we must propose a local profile for the turbulent luminosity ($L_e$). Previous simulations have suggested that it is zero at the gain radius and quickly rises to a plateau all the way to the shock \citep{murphy13}. Hence, we define the turbulent luminosity as 
\begin{equation}
L_e = L_e^{max}\tanh \left ( \frac{r-r_g}{\textbf{h}} \right ) \, ,
\end{equation}
where $\textbf{h}$ is approximately the distance it takes for $L_e$ to reach its maximum. There have been few investigations into the proper way to parameterize this value and so, as in \citet{mabanta18}, we say it scales linearly in gain region length. Various values of $\textbf{h}$ have been tested in \citet{mabanta18} showing a negligible dependence on the parameter. In \citet{mabanta18}, they simply took its value to be a third of the length of the gain region. Since our gain region is constantly evolving in time, we take the value of $\textbf{h}$ to be the minimum of $(r_{shock}-r_{gain})$/3 and 100 km. Though there is little dependence on $\textbf{h}$, as the shock proceeds to expand, this length scale approaches unrealistically large values.

Thus, we have closed the Reynolds decomposed equations with reasonable assumptions. For more detailed derivations of these equations, refer to \citet{mabanta18}. 

\subsection{Turbulence Model for PNS Convection}
\label{TMPNS}
Convection in the gain region is inefficient, in that convection
  does not completely flatten the entropy gradient.  For this reason,
  one must develop a global convective model based on energy balance.
The protoneutron star convection between radii of ~15 km and ~50 km
usually has a flat entropy gradient and is therefore efficient.
Therefore, the convective model need not be accurate, just efficient
enough to reproduce the profiles.
 To model this in our 1D+
model, we add an entropy-flattening flux for PNS
convection. The flux should be proportional to the entropy gradient
and it should transport the flux relatively quickly.  Therefore, we
propose the following simple convective flux 
\begin{equation}
    F_e = - \rho H T v_c \nabla s \, ,
\end{equation}
where $H = (d \ln \rho /dr)^{-1}$ is the density scale height,
$T$ is temperature, and
$s$ is the entropy.  $v_c$ is a pseudo convective velocity; this velocity
needs to be large enough to flatten the gradient but also satisfy
typical time-step limited stability conditions.  Therefore, we set
$v_c = .1 c_s$ where $c_s$ is the sound speed.  The resulting convective flux helps to properly transport entropy and
  flatten the entropy profile in a similar manner to the multi-dimensional cases. T

\subsection{Numerical Techniques}

To test the convective models, we include them in FORNAX, a multi-dimensional, radiation hydrodynamics, finite
volume code primarily for the use of exploring astrophysical
systems. It utilizes Runge-Kutta integration, an HLLC Riemann solver,
has potential for a non-uniform grid, and is logically Cartesian. For
an in-depth description of the code, refer to
\citet{skinner18}. The calculations of this study use either a
  spherically symmetric one-dimensional grid or an azimuthally
  symmetric two-dimensional grid. In this study, we use a simple light bulb model for neutrinos.  

The one-dimensional simulations
  have 678 radial zones, and the two-dimensional simulations have the
  same radial zoning and 256 zones in $\theta$.   The
  radial coordinate is given by $r(x) = r_t \sinh \left ( x / r_t
  \right )$ where the scale, $r_t \approx 50$ km.  This function creates a roughly
  uniform grid between $r=0$ and $r=50$ km with a resolution of $\sim$ .5 km.
Exterior to this, the grid is logarithmic and extends out to 20,000 km. 
The $\theta$ coordinate is given by 
\begin{equation}
  \vartheta (x) = \frac{\pi}{2} \left ( 1 + \frac{xB^A(A+1)+x^{A+1}}{1+B^A(A+1)}\right )
\end{equation}
 where, for the two-dimensional case, $A = 4$ and $B = 1.2$. 



The inner boundary conditions at the center and axis are reflecting.
The outer boundary follows the Dirichlet boundary condition, but since this is so
far from the shock, it is out of contact of any kind with
the shock during the time of the simulation.  The progenitor model
represents the initial conditions and is the 13 $\textrm{M}_\odot$ model of \citet{wh07}.


Rather than initiating the turbulence model from the beginning,
  we gradually ramp up the model from .2 to .25 s post bounce.  We do
this for three reasons.  For one, in multi-dimensional simulations, convection
takes time to develop in about one eddy-turn-over time (50 ms).  Two,
the current turbulence model has only been tested during the
relatively stable accretion rate phase of collapse; for our progenitor this corresponds to after about 200 ms
post bounce.  Third, the turn on is slow to avoid dramatic adjustments
in the profiles and to better mimic the gradual development of
turbulence in the multi-dimensional simulations.

Since the timescale for the eddies to reach their maximum amplitude is roughly 50 ms, we have included an amplification function, $f(t)$, which starts at 0 and linearly increases to 1 over a 50 ms interval, starting at the turn-on time. Hence, we calculate this function in the following way, 
\begin{equation}
f(t) = (t - (t_{bounce} + t_{turn-on})) / t_{ramp} \, .
\end{equation}
where $t_{ramp}$ is the ramping timescale and $t_{bounce}$ is calculated as the time it takes in the simulation for the density in the core to exceed 10$^{14}$ g cm$^{-3}$. Though this function may be negative or greater than unity, We have ensured in the code that this function is restricted to be between 0 and 1. 


\section{Results \& Discussion}
\label{results}

\begin{figure}
\epsscale{1.3}
\plotone{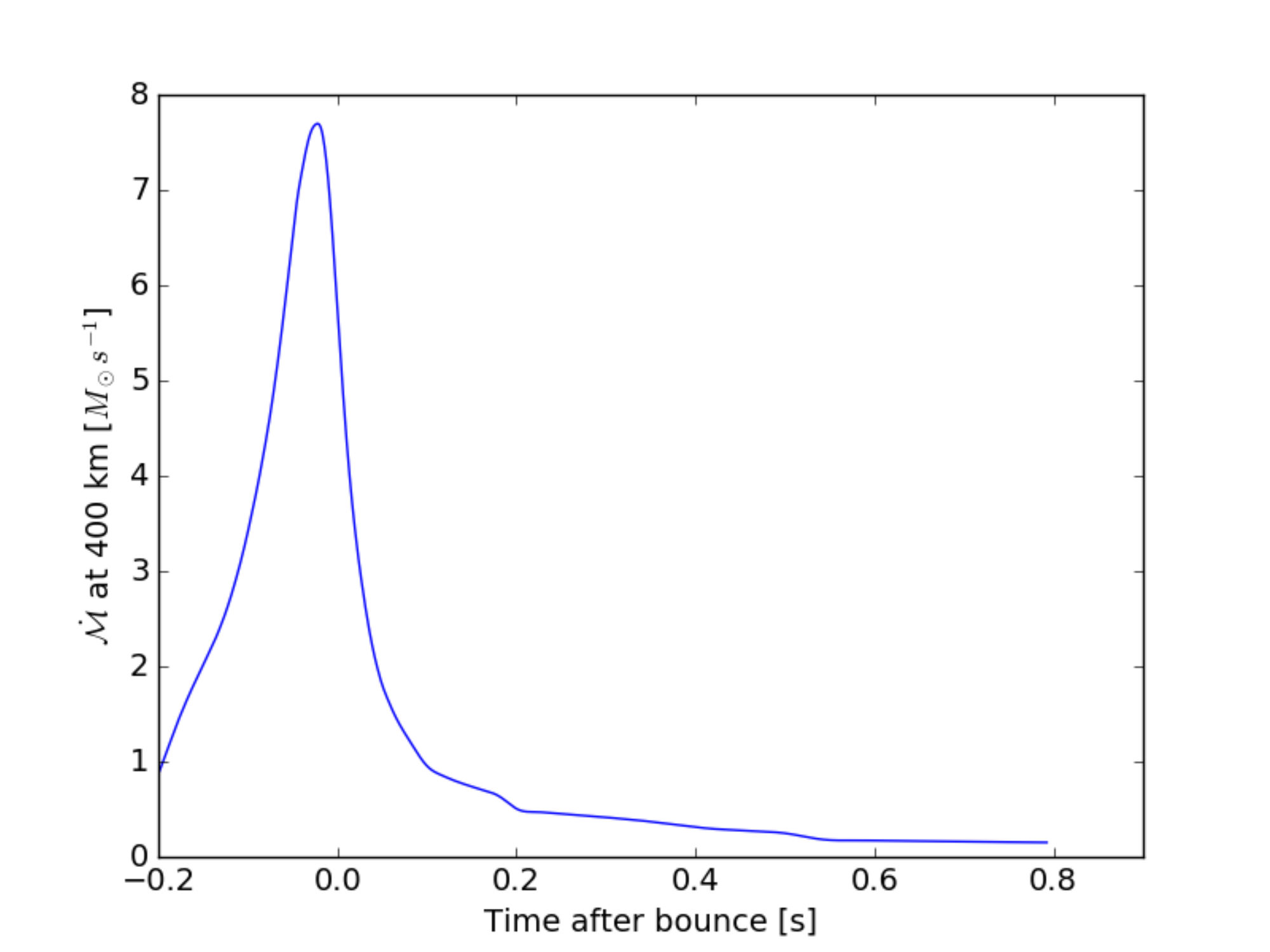}
\caption{An example mass accretion rate
  ($\mdot$)  versus time after
	bounce for the 13 M$_{\odot}$ progenitor \citep{wh07}. The accretion rate
	evolution determines the density and inward ram pressure
  at the shock and sets the outer boundary conditions of the
	post-shock structure.  Consequently, it affects the explosion  timescale. During the steady, stalled phase (after 0.1 s), the
  mass accretion rate is around 0.5 \msun\ s$^{-1}$ and slowly
  declines.  The accretion rate drops significantly at around 0.2 s
  and 0.5 s.} 
\label{mdot}
\epsscale{1.}
\end{figure}

\begin{figure}
\epsscale{1.1}
\plotone{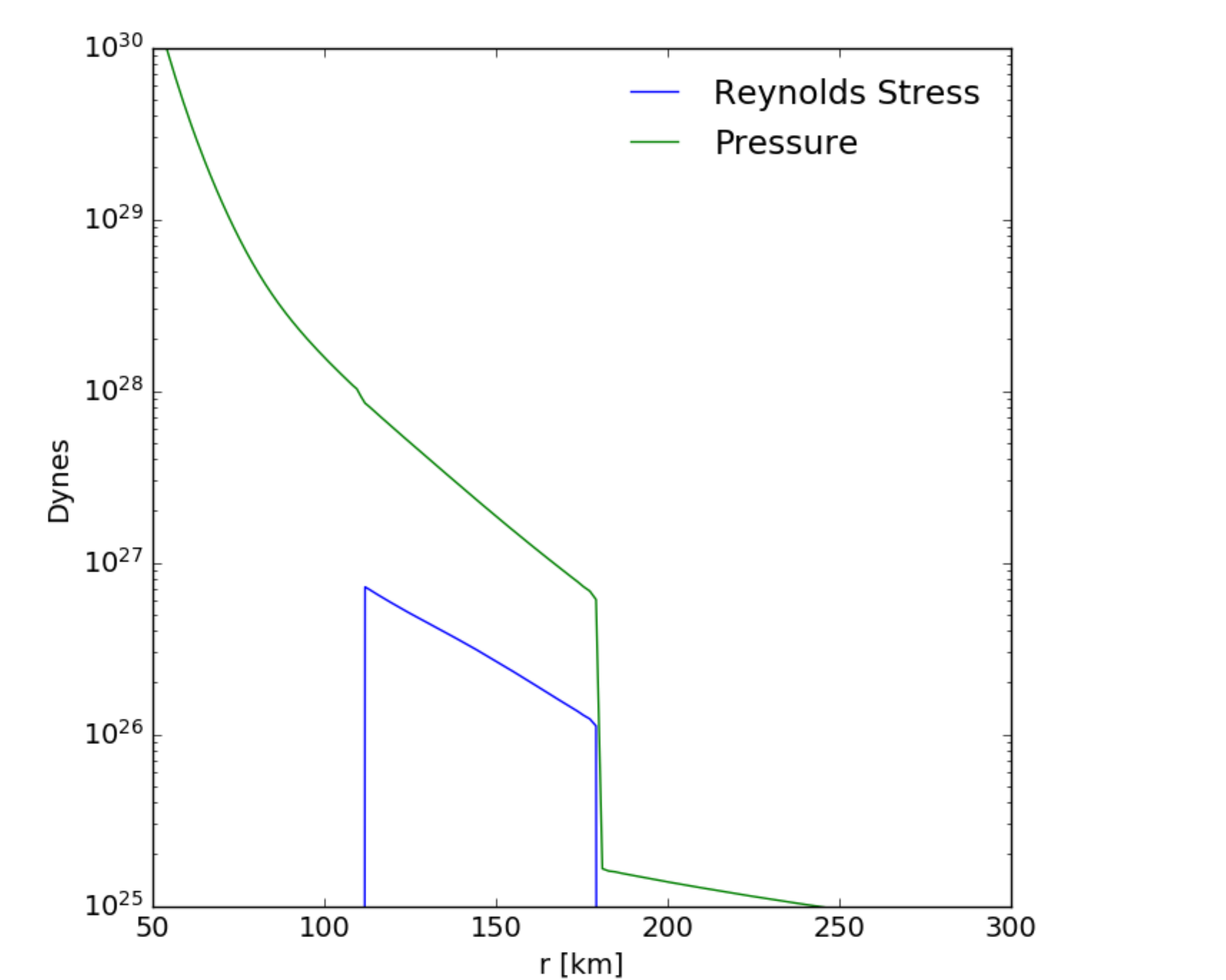}
\plotone{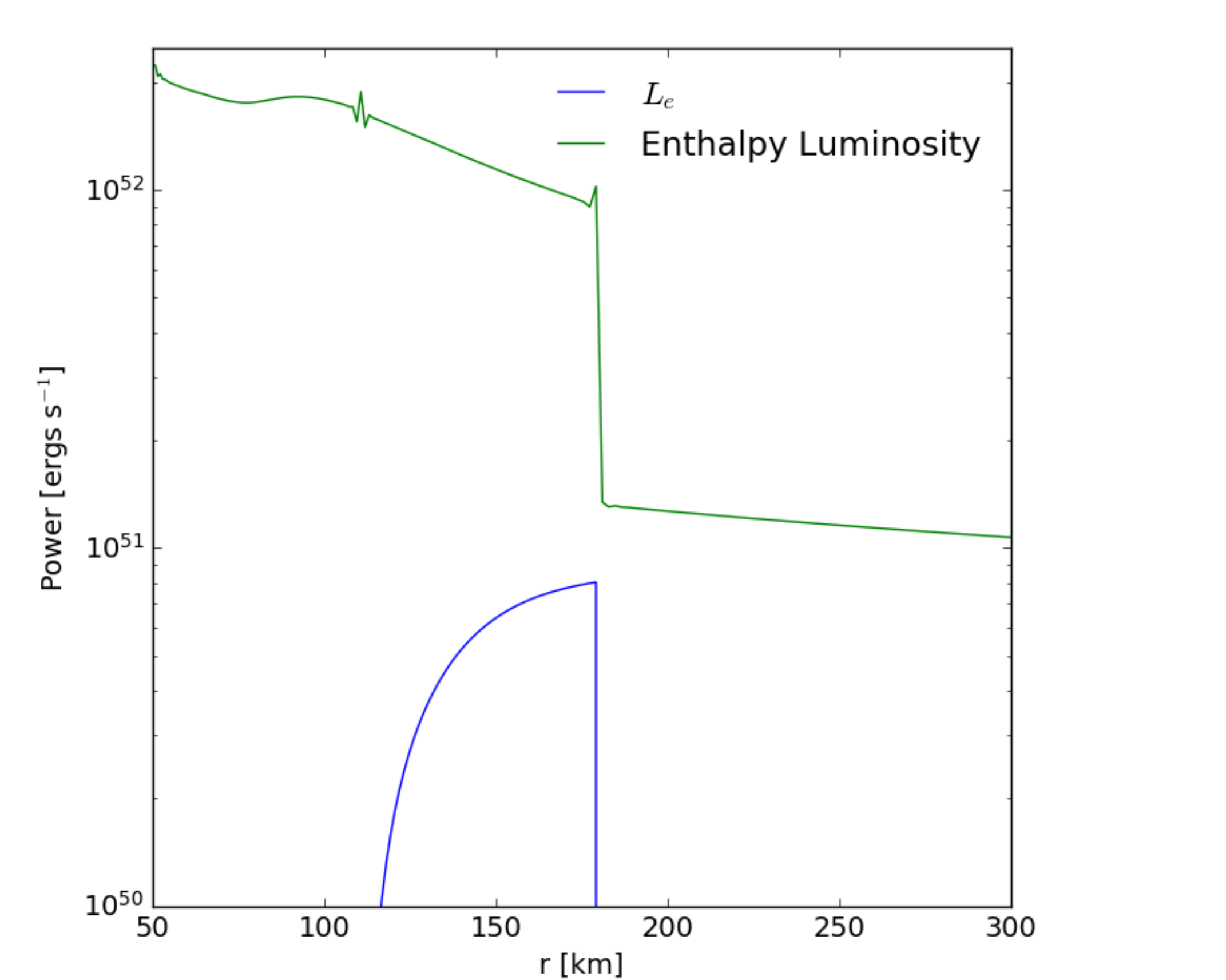}
\plotone{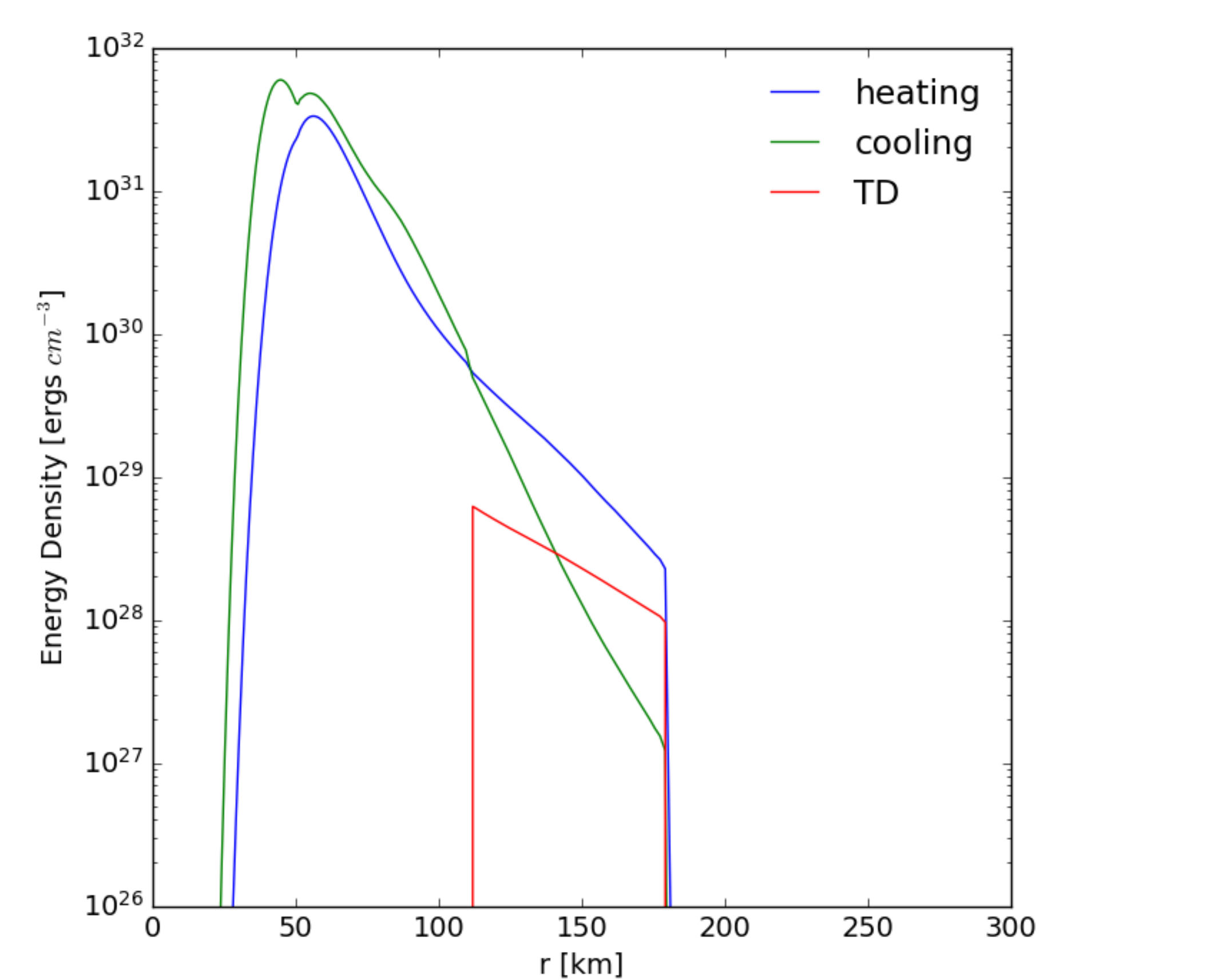}
\epsscale{1.}
\caption{Background and turbulence model profiles.  As in
  Figure~\ref{mdot}, the progenitor is the 13 M$\odot$ model of
  \citet{wh07}.  The neutrino luminosity for this simulation is
  $L_{\nu} = 2.1$ in units of $10^{52}$ erg s$^{-1}$.  The
	turbulent model is derived in \citet{mabanta18} and mimics the
	results of three-dimensional simulations \citep{murphy13}. 
The top panel compares the radial component of the Reynolds stress
($\rho R_{rr}$) with the
pressure. The middle panel compares the turbulent luminosity with the
background enthalpy flow. The bottom panel compares the $\nu$ heating
and cooling with turbulent dissipation (TD). Contrary to \citet{mabanta18}, we find that the turbulent dissipation is consistently lower than the heating. Some justifications for this discrepancy may be changes in the progenitor or other physical parameters.}
\label{comparing}
\end{figure}

\begin{figure}
\epsscale{1.2}
\plotone{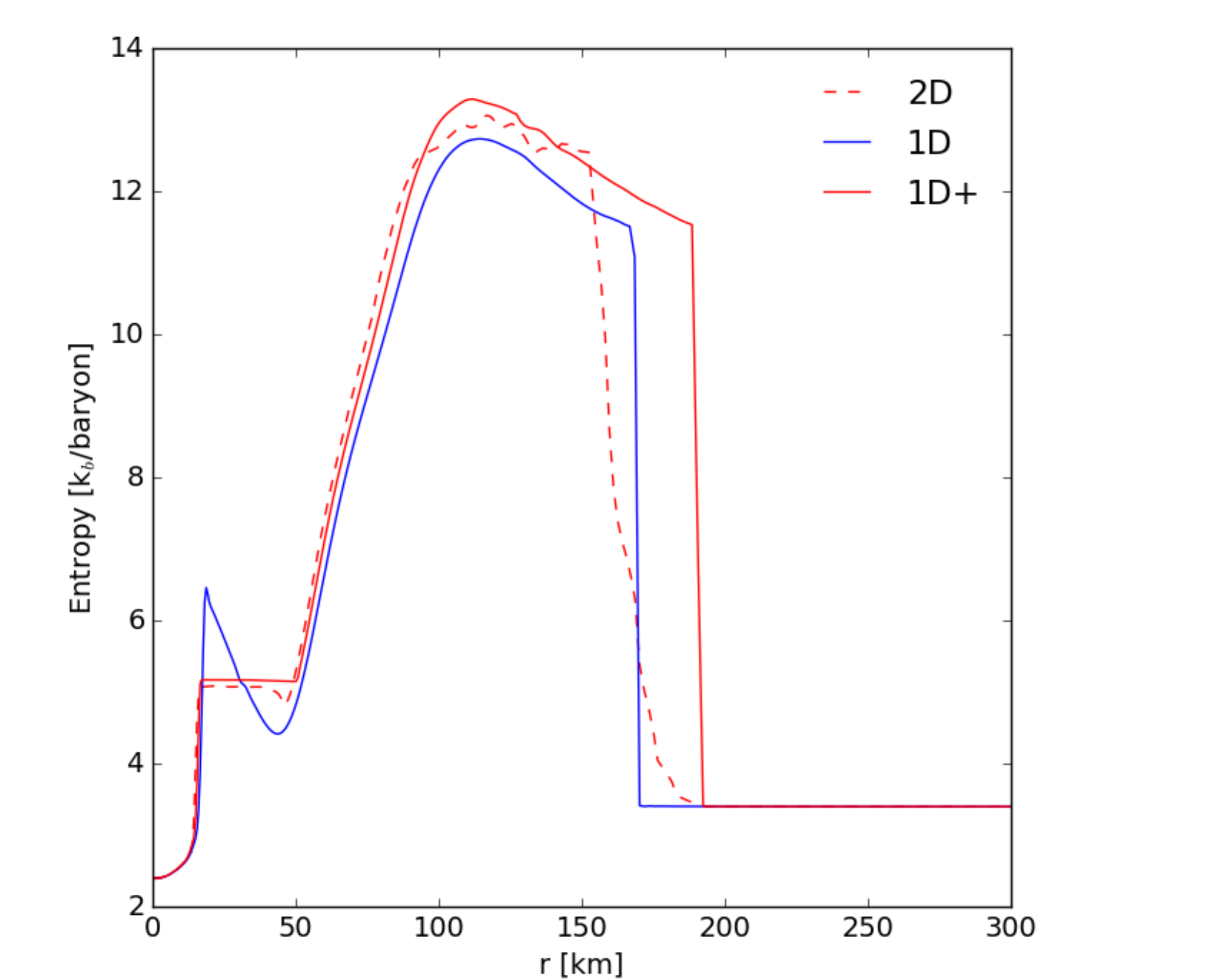}
\epsscale{1.}
\caption{Entropy profiles of a one-dimensional,
  two-dimensional, and 1D+ model. In general, the 1D+ simulation
	mimics the two-dimensional simulation and differs from the
	one-dimensional simulation in several ways. Most notably, the
	entropy is higher for both the 1D+ and two-dimensional
	simulations.  In the 1D+ simulation, this is due to the turbulent
	dissipation term, which presumably causes the higher entropy in
	the two-dimensional simulation as well.  The 1D+ shock radius is
	somewhat larger than the average shock radius for the 2D
	simulation.  This may provide a way to more accurately calibrate
	the convection model with future three-dimensional simulations.}
\label{entropy}
\end{figure}

\begin{figure}
\epsscale{1.2}
\plotone{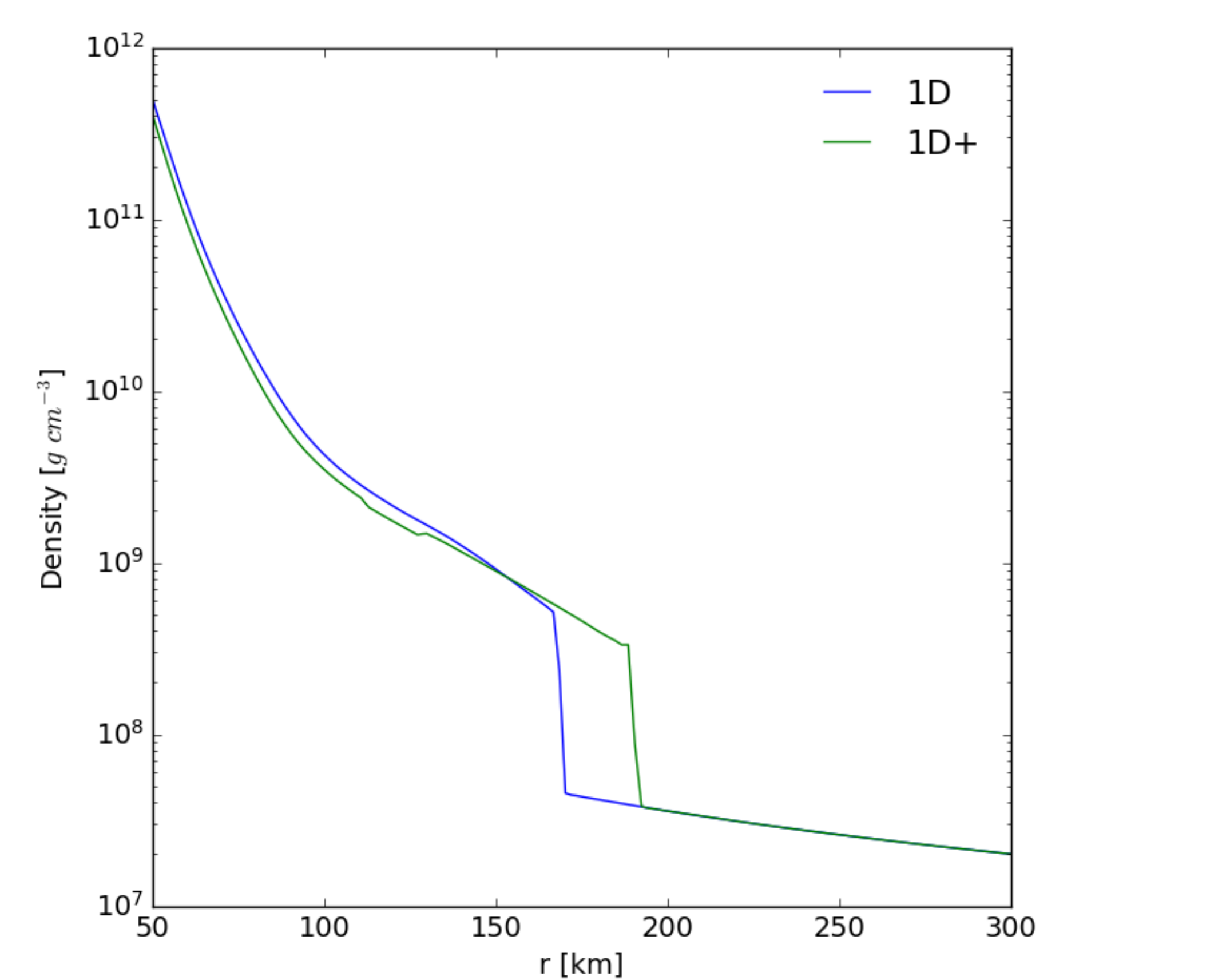}
\plotone{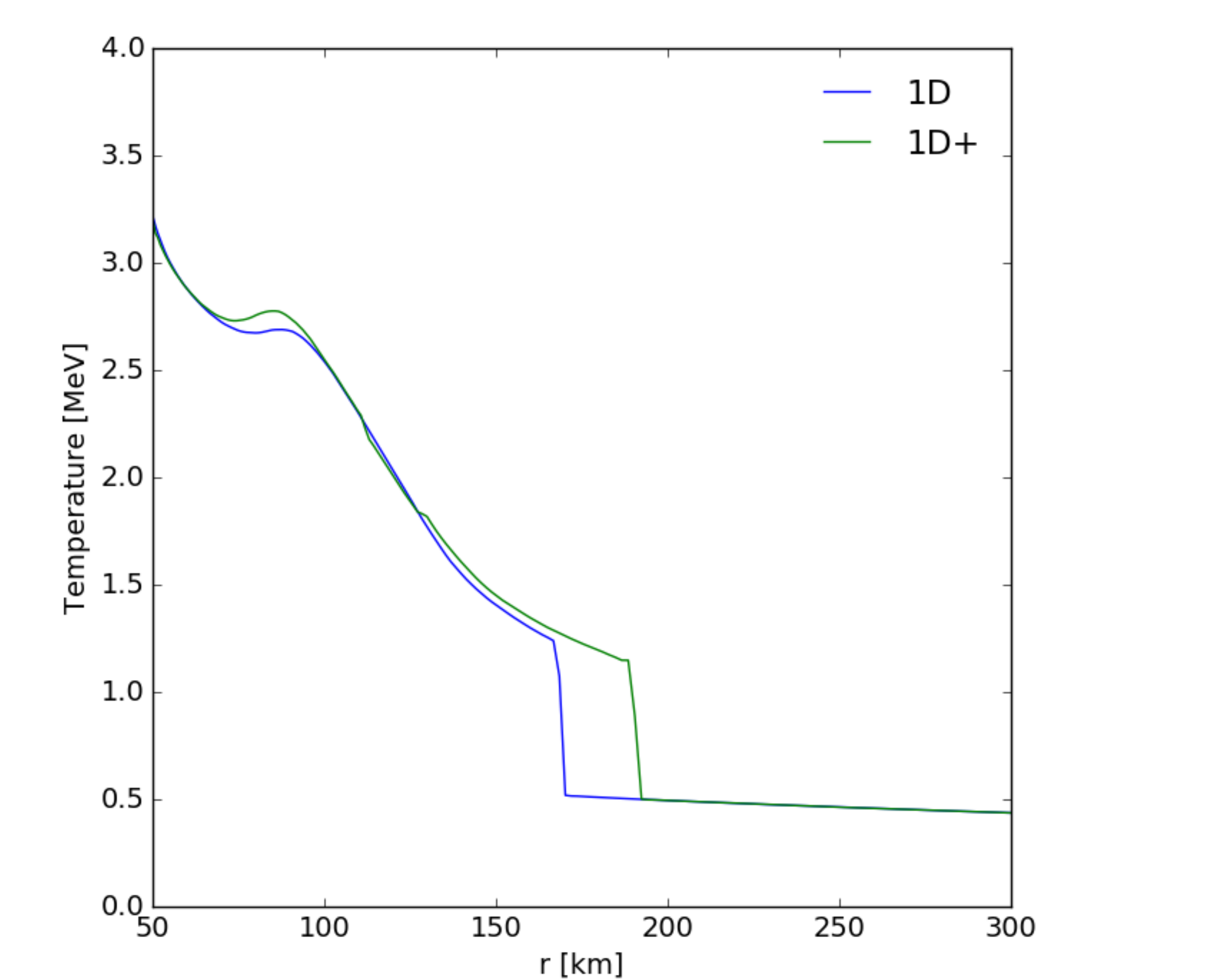}
\epsscale{1.}
\caption{ One-dimensional density and temperature profiles with and
  without the turbulence model. Though
  the changes are subtle, two important differences are the farther
  shock radius in the 1D+ case, and the shallower density
  gradient. \citet{murphy17, mabanta18} note that a shallower post-shock density gradient leads to an ease of explosion.}
\label{denstemp}
\end{figure}

\begin{figure}
\epsscale{1.2}
\plotone{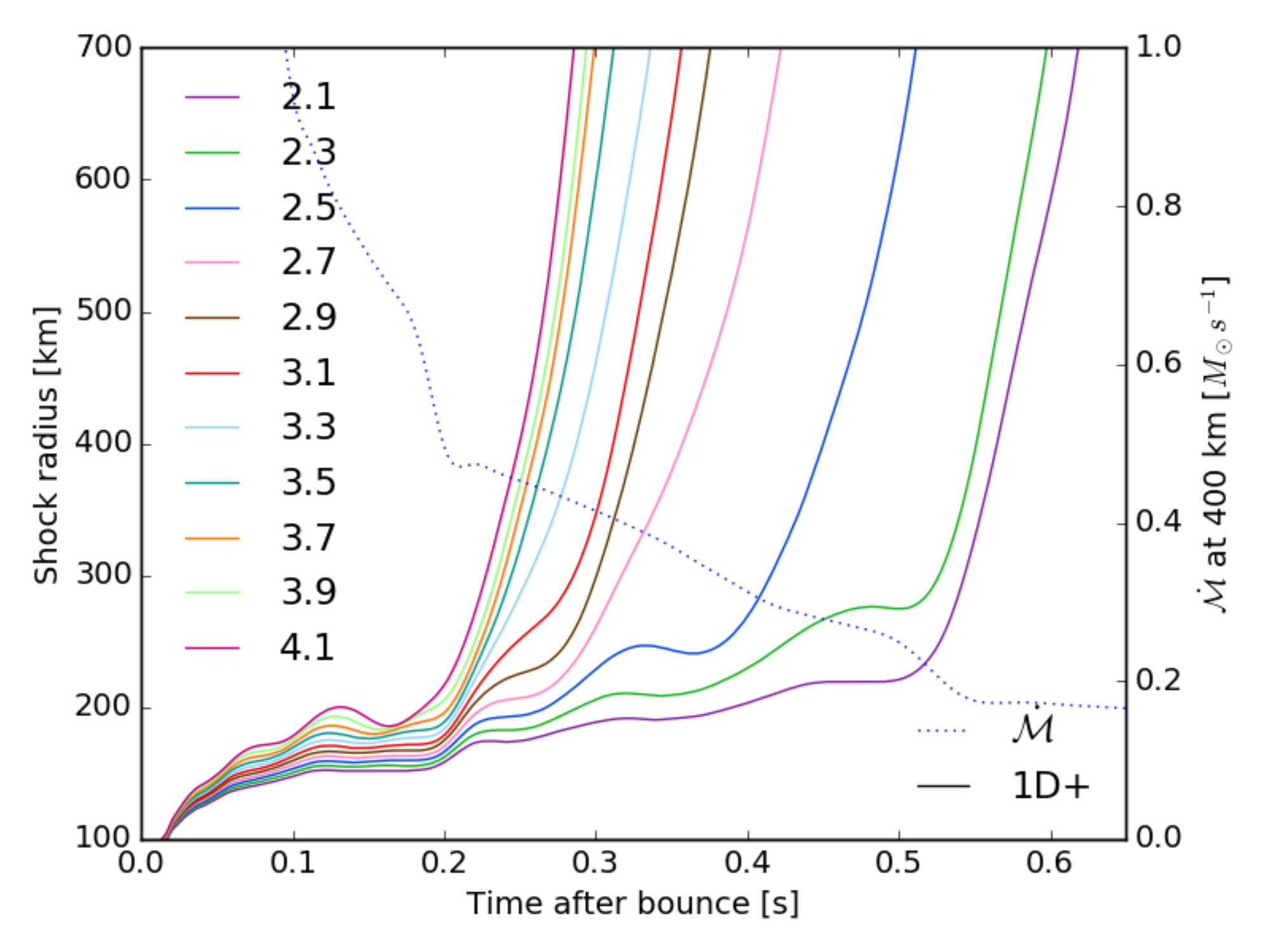}
\caption{Shock radius vs. time after bounce for the 1D+ simulations.
  Each curve is labeled by the neutrino luminosity in units of
  $10^{52}$ erg s$^{-1}$.
For reference, the dotted-blue line and the right
	horizontal axis show the mass accretion rate 
  ($\mdot$). For each model, we note the time of explosion, the
	corresponding accretion rate, and mark the point (L$_\nu$,$\mdot$)
  in the critical curve of Figure~\ref{lcrit}.  We define
  explosion as the point where the shock exceeds 400 km}
\label{expcond}
\epsscale{1.}
\end{figure}

\begin{figure}
\epsscale{1.2}
\plotone{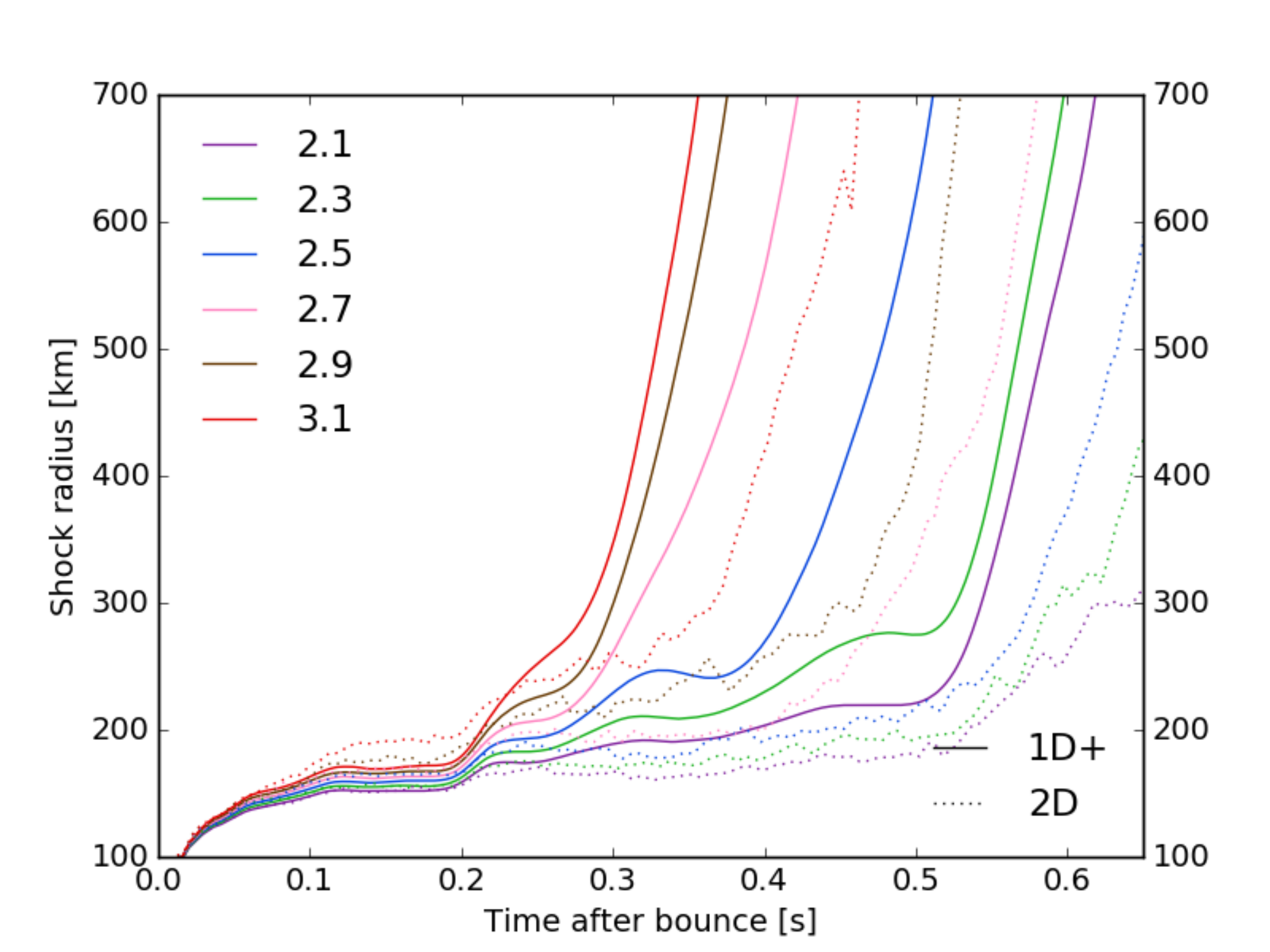}
\caption{Shock radius vs.\ time after bounce for two-dimensional
	and 1D+ simulations.  Again, the label indicates the neutrino
	luminosity in units of $10^{52}$ erg s$^{-1}$.
For the most part, the 1D+ and 2D models explode at similar times;
	the 1D+ models explode slightly earlier, but this makes little difference in
  the critical curves in Figure~\ref{lcrit}.  To obtain better
	agreement, one might further tune the turbulence model. }
\label{MDrshocks}
\epsscale{1.}
\end{figure}

\begin{figure*}
\epsscale{1.2}
\plotone{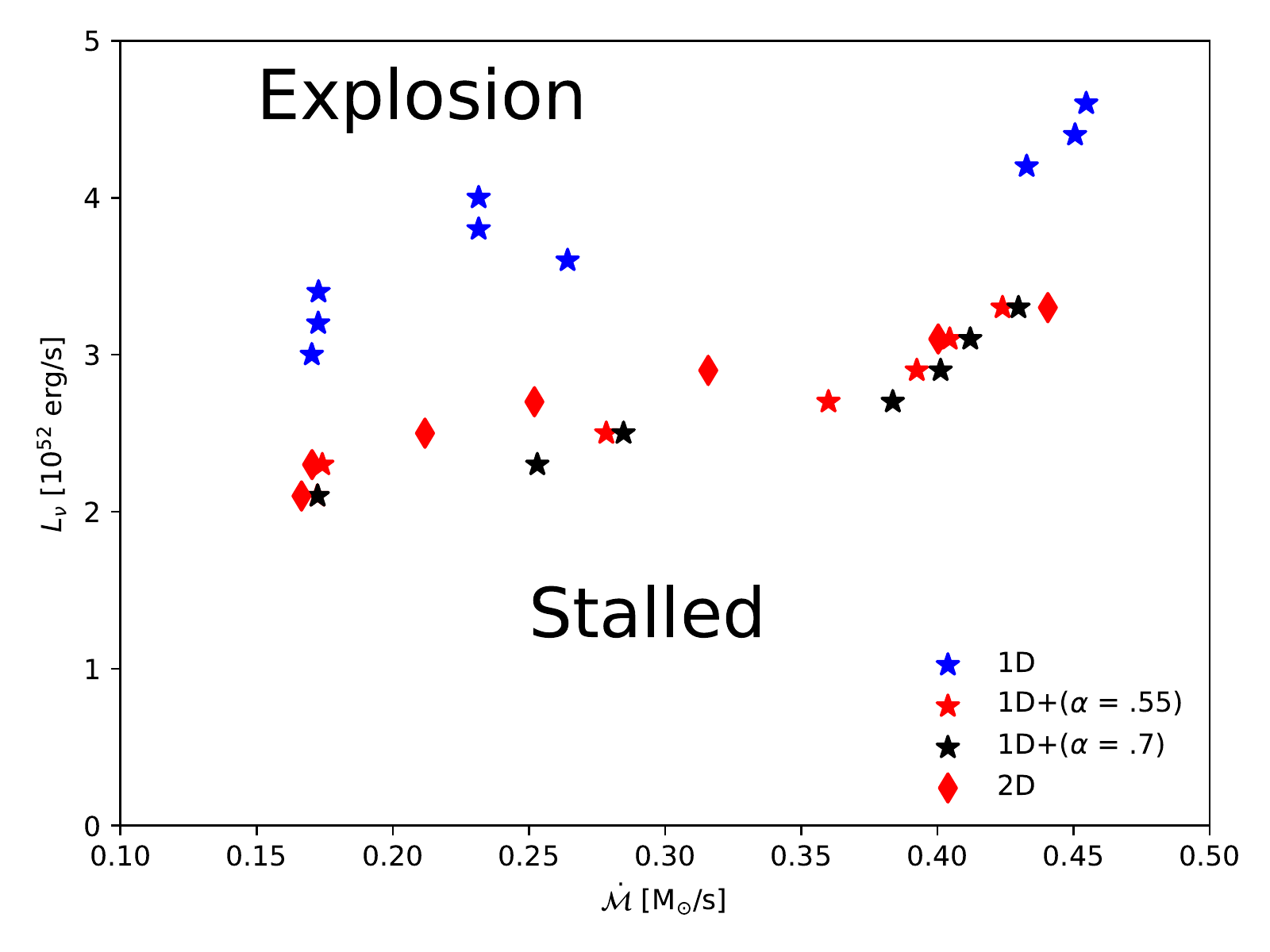}
\caption{An empirical slice of the condition for explosion: the
  neutrino-luminosity and accretion-rate critical curve.  As has been noted in other simulations, the two-dimensional case requires $\sim$
  30\% less neutrino luminosity to explode than the one-dimensional
  case.  The 1D+ critical curve mimics the explosion conditions of 2D simulations.
 The convection model has two parameters;
	  $\alpha$ relates the neutrino power to turbulent dissipation,
	  and $\beta$ relates the neutrino power to the turbulent
	  luminosity.  Two- and three-dimensional simulations show that
	  $\beta = 0.3$, but $\alpha$ is different: $\alpha = 0.7$ for 3D,
	  and $\alpha = 0.55$ for 2D.  We test the explodability for
	both $\alpha$ values
	and found little difference. For some cases, the turbulence model in this 1D+
	simulation is slightly more explosive than the two-dimensional
	case.  Further calibration of the turbulence model might
	  reduce this slight discrepancy.}
\label{lcrit}
\epsscale{1.}
\end{figure*}

The primary goal is to develop a turbulence model that enables
  one-dimensional simulations to mimic the evolution, profiles, and
  explosion conditions of multi-dimensional simulations.  First,  Figure~\ref{mdot}
  plots the mass accretion rate vs. time after bounce.
Figures~\ref{comparing}-\ref{denstemp}  show the radial profiles,
  Figures~\ref{expcond}~\&~\ref{MDrshocks}  compare the evolution between
  1D+ and 2D, and Figure~\ref{lcrit} compares
  the explosion condition. In this section, we  also discuss the implications for future projects.

  The mass accretion rate plotted in Figure~\ref{mdot} is an
  important parameter of the problem, establishing the ram pressure at
  the shock and the location of the shock.  Two important regions to note are roughly at 0.2 s
  and 0.5 s. The accretion rate drops dramatically due to significant entropy changes in the
  progenitor's profile at the boundaries of prior shell burning
  phases.  In  general, the mass accretion rate is similar for all
  progenitors, but the details differ.  See \cite{vart18} and \citet{ott18} for example
	profiles.  The magnitude of accretion rate will differ,
  and the timing of the drops will differ (c.f. \ref{expcond} for a zoom-in of this region).  The
  structure of this profile is unique to this specific progenitor, and
 the explosion outcome is sensitive to this structure.
  Hence, a primary motivation for the
  development of the 1D+ technique is to be able to
  probe the explodability of several progenitors with expedited
  numerics. 
  
 Figure~\ref{comparing} compares profiles of
  the turbulent correlations with the background.   The top
  panel of Figure~\ref{comparing}  shows that the
  Reynolds stress is a significant fraction (15\%) of the pressure within the
  gain region. Similarly, the turbulent
	luminosity (middle panel) is roughly 10\% of the enthalpy luminosity at the shock.
  The bottom panel compares turbulent dissipation
	with neutrino heating and cooling.  The ratio of turbulent heating
  to neutrino heating varies from 10\% near the gain radius to
  50\% near the shock.   Note, that the turbulent
	dissipation plotted in Figure~\ref{comparing} is $\rho \epsilon$,
	where $\epsilon = R_{rr}^{3/2}/\mathcal{L}$.  Even though we assume
	$R_{rr}$ and $epsilon$ to be constant with radius, the $\rho
	\epsilon$ is not constant only because of the density profile. The
	combined effects of turbulent ram pressure, turbulent transport,
	and, most importantly, turbulent dissipation are crucial in achieving explosion with similar conditions as
	multi-dimensional simulations.\footnote{See \citet{mabanta18}
	  for an exploration of the importance of the turbulent terms.}

 Figures~\ref{entropy}-\ref{MDrshocks}, show how
 these turbulent terms affect the
 radial profiles.  Figure~\ref{entropy} shows  the
   entropy profiles at 350 ms after bounce for the 1D, 1D+, and 2D
   simulations.   In general, the 1D+ simulation
	mimics the two-dimensional simulation and differs from the
	one-dimensional simulation in a couple of ways.  
  First, the peak entropy is higher for both the two-dimensional and
  1D+ cases.  In the 1D+ simulation, this higher entropy is due to turbulent dissipation,
  and presumably this is the case for the two-dimensional simulation.
  Second, both the two-dimensional and the 1D+ models show similar
  entropy profiles in the inner convective region
   between 15 and 55 km. As designed (see eq.~(\ref{TMPNS}), the entropy gradient is flat for the inner
	convection.  One difference is that the average shock radius of
	the 1D+ simulation has a larger radius than the 1D or 2D
	simulation.  The lack of a well-defined shock
  in the two-dimensional case is a result of an angle-averaged entropy
  profile; since the shock is not spherically
  symmetric, the shock is at different radii for different angles.
  The largest shock radii in the 2D case are similar to the 1D+ shock radius.

Figure~\ref{denstemp} shows the resulting changes in the density
  (top panel) and temperature (bottom) profiles at 350 ms post bounce.
  Both the density and
 temperature profiles are shallower as a result of convection
 (figure~\ref{denstemp}). This is consistent with the differences
 seen in steady-state explorations \citep{mabanta18}.  
 
 
 
 

 Figure~\ref{expcond} 
illustrates the shock radius evolution  for various neutrino luminosities.
For perspective, the blue-dotted line shows the corresponding
  evolution of $\mdot$ at 400 km, since we declare explosion once the shock exceeds 400
km. This is far enough from the range of oscillations, and hence,
explosion is unambiguous. Though this is an arbitrary point, this range is a
common, fiducial criterion for explosion (e.g. \citet{sukhbold16} uses 500 km). In fact, for these simplified models, past 400km,
  the shock radius never returns.  To determine the $\mdot$ associated with
explosion, one can compare the time at which a specific curve crosses
400 km and match that point to its associated $\mdot$
value. Note that many of the
high luminosity runs explode at or just after the significant drop at
$\sim$0.2 s.  This pile-up of explosion times and nearly vertical $\mdot$ vs.
time profile makes it difficult to assess a unique explosion time or
$\mdot$.  For clear comparisons, subsequent analyses focus on the range of
luminosities that give explosion after this drop.   Finding an
(L$_\nu$,$\mdot$) data point where there is a smooth mass accretion
rate curve more accurately describes the explosion
condition.  Figure~\ref{MDrshocks} shows
a subset of these shock radii for the 1D+ runs, but
  this time, the figure also includes
shock radii of the two-dimensional simulations. The higher luminosities in this plot have been pruned for clarity. These results show that the 1D+ model
explodes at similar times, but slightly earlier than their
two-dimensional counterparts.  

Figure~\ref{lcrit} compares the critical luminosities for
  one-dimensional, 1D+, and two-dimensional simulations.  In general,
  the explosion condition depends upon $L_{\nu}$, $\mdot$,
  $R_{\text{NS}}$, and $M_{\text{NS}}$ \citep{murphy17}.  In these simplified models,
  it is most straightforward to note the luminosity and accretion rate at
explosion, so we merely show the $L_{\nu}$-$\mdot$ slice of the
critical condition. Here, the blue stars
  represent explosions in one dimension, the red
diamonds represent explosions in two dimensions, and the
red and black stars are explosions in our one-dimensional convective model with $\alpha$ = .55 and $\alpha$ = .7, respectively. Note that the reduction
  between 1D and multi-D (2D in this case) is about 30\%, which agrees
with the results of other simulations \citep{murphy08b,hanke12,couch13}.  The
1D+ simulations mimic the critical condition of the 2D simulations,
but the 1D+ simulations used a factor of 100 fewer computing resources.
 If 1D+ simulations also reproduce the explosion conditions of
  three-dimensional simulations, then the 1D+ simulations could use a
  factor 100,000 fewer computing resources to predict which stars will
  explode.

\section{Conclusion}
\label{conclusion}

In general, spherically symmetric CCSN simulations often fail to explode
  while multi-dimensional simulations sometimes do explode.  However, these
  multi-dimensional simulations are very computationally expensive.
  For example, a recent three-dimensional simulation which ended in a
  successful explosion required 18 million CPU-hours on 16,000 cores;
  this one simulation took 1.5 months to compute, and it is one of the
  most efficient 3D simulations to date \citep{vart18}.
  \citet{sukhbold18} noted that the progenitor structure shows
  significant variation for a small range of masses, and this was
  for one set of stellar evolution parameters.  It may require thousands of CCSN simulations to
  predict which stars will explode and which won't.  At this rate, it
  would take hundreds of years to simulate enough progenitors to produce a statistically significant data set.
  We present a method to incorporate  neutrino-driven convection into one-dimensional simulations.  These
  augmented simulations are called 1D+, and they mostly reproduce the
  profiles and explosion conditions of simple two-dimensional simulations.

The 1D+ simulations include a neutrino-driven convection model that
  was derived using Reynolds Decomposition
  \citep{murphy11,murphy13,mabanta18}.   It includes Reynolds stress
  in the momentum equation, turbulent flux in the energy equation, and
  turbulent dissipation in the energy equation. Two and three dimensional models were used to 
  calibrate the neutrino-driven convection model using
  three-dimensional simulations with simple neutrino heating and
  cooling \citep{murphy13}.  \citet{mabanta18}
  included these terms in the steady-state equations describing the
  stalled shock.  They found that the turbulence reproduces the
  reduction in the critical condition for explosion seen in
  multi-dimensional simulations.  What is more, they found that
  turbulent dissipation is a dominant effect in this reduction.  The
  1D+ simulations of this manuscript uses the same neutrino-driven
  convection model, but here we reformulate the model for
  time-dependent one-dimensional simulations.  These particular 1D+
  simulations use a simple neutrino heating and cooling prescription,
  but the model may work just as well for more sophisticated neutrino
transport.

Qualitatively, the 1D+ model reproduces the radial profiles and
  explosion conditions of simple two-dimensional models.
  Furthermore, we show how the turbulent terms influence the radial profiles.   We have also included the evolution of shock radii for both our 1D+ model and the two-dimensional runs at relevant luminosities. Lastly, we have shown that this technique accurately emulates the luminosity reduction for a successful explosion seen previously in multi-dimensional simulations.

Though 1D+ successfully mimics the explosion condition and profiles
of multi-dimensional simulations before explosion, there are several details that the
1D+ model may not replicate. For one, the explosion seems to be
  inherently aspherical.  Simulators noticed that 3-dimensional
  simulations seem to be dominated by one large buoyant plume \citep{dolence13,lentz15,muller18,vart18}.  So,
  while a spherically averaged mean-field convection
model seems to reproduce the multidimensional instabilities before
explosion, it may not be possible to mimic how explosion develops in a
1D+ model.  If true, with the current version of 1D+, this would limit the reliability of post
explosion diagnostics such as explosion energy, neutron star masses,
nucleosynthetic yields, etc. Going forward, we will compare these
post explosion diagnostics with three-dimensional simulations; one may
be able to develop mean field models that reflect the
multi-dimensional post explosion instabilities.  However, since 1D+ is
only calibrated to reproduce the explosion conditions of
multi-dimensional simulations, our primary focus will be to
use 1D+ to predict which stars explode by the neutrino and convection
mechanism.

Thus far, the current implementation of 1D+ is consistent with
  two-dimensional simulations using simple neutrino heating
  and cooling and Newtonian gravity.  The eventual goal is to
  reproduce the explosion conditions of relativistic three-dimensional
  radiation-hydrodynamic simulations of CCSNe.  The current turbulence
  model has  been calibrated for only a handful of two- and
	three-dimensional simulations \citep{murphy11,murphy13}.  More
	simulations are necessary to validate the
	turbulence model in the wide range of conditions seen in core
	collapse simulations.  For example, the neutrino transport of this
	manuscript is too simple for predictions of explosion; such
	simulations will require validation with a more nuanced neutrino
	transport scheme such as two-moment closures \citep{roberts16,just18,oconnor18,skinner18,vart18}.
	Furthermore, the turbulence model has only been calibrated with
	Newtonian monopole gravity.  A full validation will require
	comparisons using non-spherical general relativity.  Finally, the
	turbulence model and explosion conditions have only been validated
	using a limited set of progenitors.  A study considering a wider
	range of structures and accretion rate histories would help to
	validate the 1D+ algorithm in the full context of core collapse.  

\section{Acknowledgments}
Quintin A.~Mabanta and Joshua C.~Dolence acknowledge support from the Laboratory Directed Research and Development Program at Los Alamos National Laboratory.  This research used resources provided by the Los Alamos National Laboratory Institutional Computing Program, which is supported by the U.S. Department of Energy National Nuclear Security Administration under Contract No. 89233218CNA000001. This work has been assigned an LANL document release number LA-UR-19-20695. Support for this work was also in part funded by the National Science Foundation under Grant No. 1313036.

\end{document}